\documentclass[twocolumn,showpacs,preprintnumbers,amsmath,amssymb,aps,pra]{revtex4}

\usepackage[dvips]{graphicx}
\usepackage{dcolumn}
\usepackage{bm}
\usepackage[mathcal]{euscript}

\def\beq{\begin{equation}}
\def\eeq{\end{equation}}

\begin{document}

\preprint{}

\title{Robust quantum searching with spontaneously decaying qubits}

\author{Robert J.\ C.\ Spreeuw}
\email{spreeuw@science.uva.nl}
\author{Tom W.\ Hijmans}
\affiliation{Van der Waals-Zeeman Institute, University of Amsterdam, \\
Valckenierstraat 65, 1018 XE Amsterdam, The Netherlands}

\date{\today}

\begin{abstract}

We present a modification of the standard single-item quantum search procedure 
that acquires robustness from spontaneous decay of the qubits. 
This damps the usual oscillation of populations,  
driving the system to a steady state with a strongly enhanced population 
of the solution. 
Numerical evaluation of the steady state was performed for up to 36 qubits. 
The huge size of the state
space in our analysis is dealt with by exploiting a symmetry in the master 
equation that reduces the scaling of computer resources
from exponential to polynomial. Based on these results we estimate that
an error-free solution can be retrieved from the steady state after $O(\log \log N)$ repetitions, 
with near-unit probability. This brings the overall scaling to $O(\sqrt{N}\,\log \log N)$, only 
slightly worse than for the ideal quantum case. 

\end{abstract}



\maketitle

\section{Introduction}

The major challenge in quantum information science and technology
is to achieve complete control over quantum systems. In this light it has
been realized early on that loss of coherence can be a prohibitive
problem for quantum computers \cite{Q:Unr95,Q:PalSuoEke96,Q:MiqPazPer96}. 
In general, spontaneous and incoherent processes are therefore best avoided 
where possible, or counteracted by quantum error 
correction \cite{Q:Sho95,Q:Ste96b}. From this perspective it is 
remarkable that in specific cases dissipation can also be put to use. 
Earlier work has shown that dissipation and noise can assist in the production of, or even 
generate entanglement \cite{Q:PleHueKni99,Q:BeiBosVed00,Q:PleHue02}. 
Here we modify Grover's quantum search algorithm \cite{Q:Gro97} to yield a different, 
more robust, operating procedure in the presence of local decay.
Instead of applying the search iterations a
prescribed number of times, we let the
system relax to a steady state by spontaneous decay.

Several authors have investigated the effects of errors on the performance of
quantum algorithms.
Most studies have concentrated on unitary errors, including random errors (noise)
or systematic (static) imperfections
\cite{Q:MiqPazZur97,Q:PabRui99,Q:LonLiTu00,Q:SonKim03,Q:Bet04}.
In this paper we study errors of a nonunitary, dissipative nature. These have
previously been investigated in various different contexts
\cite{Q:Unr95,Q:PalSuoEke96,Q:MiqPazPer96,Q:ZhiShe06}.
Experimentally, the effects of losses were studied in an optical, classical-wave analog
of quantum searching \cite{Q:HijHuuSpr07,Q:BhaLinSpr02}.

Robustness against errors and resistance to decoherence has previously been reported for 
adiabatic quantum computation \cite{Q:ChiFarPre01}. Although aiming for similar benefits, 
a few essential differences with our approach presented here are worth pointing out. 
Adiabatic quantum computation requires dynamic control over the Hamiltonian as 
the computation progresses. In our approach the dynamic control is entirely absent, 
to the extent that the initialization as well as the timing of the computation are rendered
superfluous. 
A second essential difference concerns the role of the dissipation. Whereas adiabatic quantum 
computation has been reported to be robust against decoherence, 
in our modified quantum search procedure a specific type of dissipation 
plays an {\em active role}: it is used to drive the 
system to a steady state. Consequently, the acquired robustness is of a different nature.

The idea of the present paper can be understood as follows.
The quantum search algorithm essentially drives a rotation in a two-dimensional subspace of the full
Hilbert space, spanned by the initial state and the solution. Noise and incoherent processes
will usually lead to loss of performance by ``leakage'' of population out of this subspace \cite{Q:SonKim03}.
We show that we can prevent the system from straying too far from this 
two-dimensional subspace, if the noise is of a ``natural'' origin: spontaneous qubit decay. 

We numerically simulate a single-item quantum search for a marked solution, in the 
presence of spontaneous decay, on our classical computer. 
In principle we would suffer from an exponential
scaling of resources \cite{Q:Fey82}. Fortunately, for the
problem at hand we identify a symmetry that reduces the scaling to polynomial
in the number of qubits, with $\sim q^3$ density matrix elements. 
By exploiting this symmetry we can handle  up to
36 qubits on a desktop personal computer.

\begin{figure}[b]
\includegraphics[width=65mm]{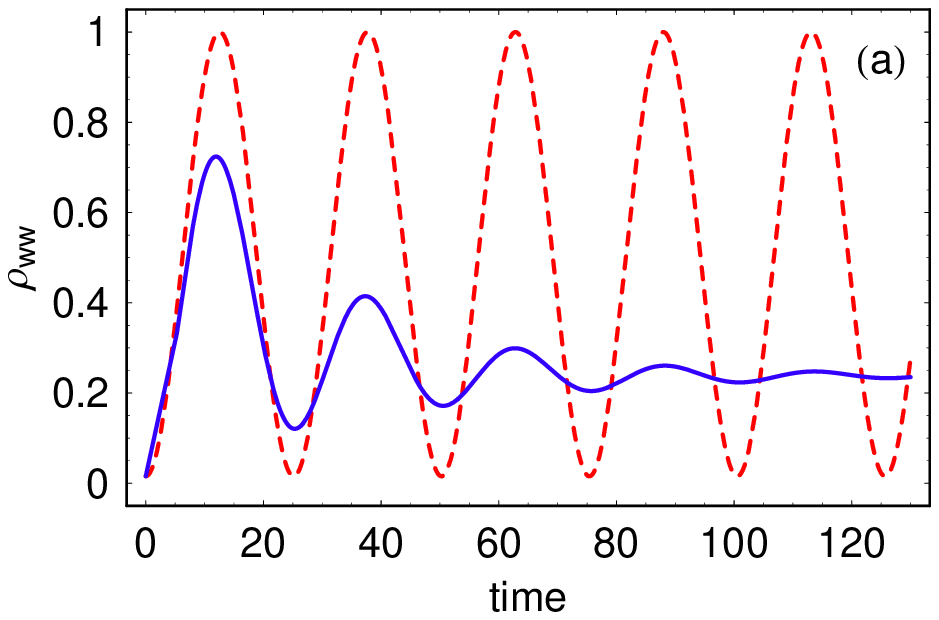}

\bigskip

\includegraphics[width=65mm]{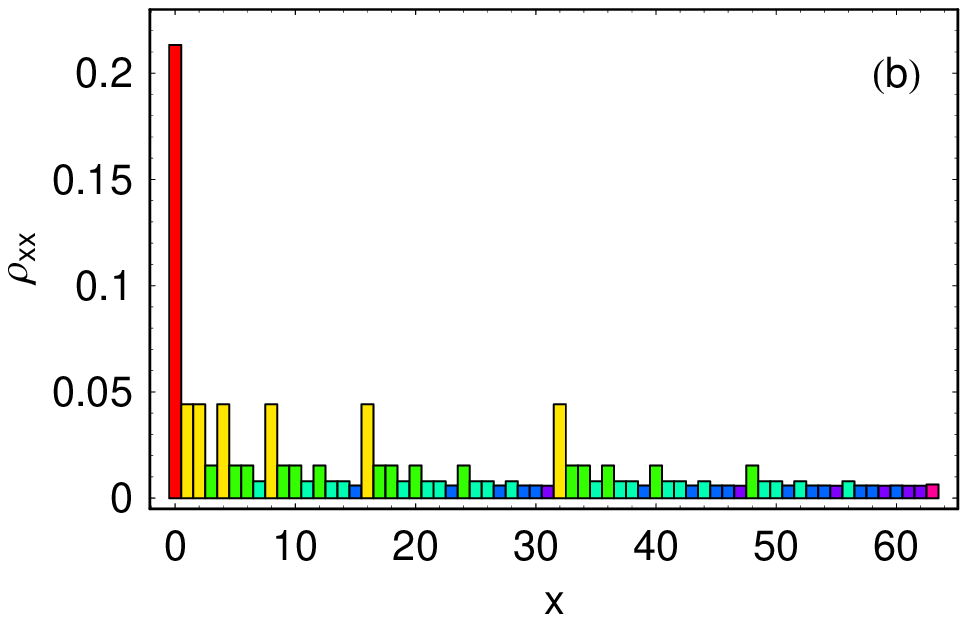}
\caption{(Color online) (a) Evolution of the population of the solution $\rho_{ww}$
without decay (red dashed line) and with $\Gamma =0.03$ (blue solid line),
for $q=6$ qubits.
(b) Steady state populations $\rho_{xx}$ for all basis states for the above case
with decay. The solution is $x=0$; equal colors indicate equal Hamming distance 
to the solution.}
\label{F:dampos}
\end{figure}

The remainder of the paper is structured as follows. In Sec.~\ref{sec:formalism} we adopt
the Hamiltonian description of a single-item quantum search and we extend this 
description to include also decay processes. We give the resulting master equation for the 
density matrix that includes our specific choice for the decay model, 
which is crucial for the results in this paper. 
In Sec.~\ref{sec:symmetry} we identify a symmetry in the steady state that allows us to 
drastically reduce the amount of data in a reduced density matrix. In Sec.~\ref{sec:results}
we discuss the numerical results obtained by solving for the steady state. 
We discuss how one would obtain an error-free solution by repetition and majority voting.

\section{Formal description}
\label{sec:formalism}

\subsection{Search Hamiltonian}

In the usual formulation of Grover's quantum search algorithm \cite{Q:Gro97}, one first
initializes a quantum register with all qubits in the state $|0\rangle$,
and applies a bitwise Hadamard operation $H^{\otimes q}$. Throughout this paper 
we indicate the number of qubits by $q$. The bitwise Hadamard yields a
symmetric superposition of all basis states, $(H |0\rangle)^{\otimes
q}$, which we write as
\beq
|s\rangle=(H |0\rangle)^{\otimes q}=2^{-q/2}\sum_{x=0}^{2^q-1}|x\rangle.
\label{eq:zerotilde}
\eeq
To this state one then repeatedly applies a unitary search iterator 
$G_w=(2|s\rangle\langle s|-I)(I-2|w\rangle\langle w|)$, 
where $w$ is the solution, and $I$ the identity operation. 
As a result, the population of the solution $\rho_{ww}$ 
oscillates between 0 and 1. 
The readout is usually performed after $(\pi /4)\sqrt{N}=(\pi /4)2^{q/2}$ 
iterations, when $\rho_{ww}\approx 1$.

In this paper we shall use an alternative formulation in terms of time-continuous 
evolution $\dot{\psi}=-i\mathcal{H}_w\psi$, with the search 
Hamiltonian given by 
\beq
\mathcal{H}_w = |w\rangle\langle w| + |s\rangle\langle s|.
\eeq
This formulation, which has been introduced by Farhi and Gutmann \cite{Q:FarGut98},
yields very similar behavior, with an oscillation period of $\pi\sqrt{N}=\pi\,2^{q/2}$.
The Hamiltonian formulation has previously been applied to discuss adiabatic quantum searching 
\cite{Q:RolCer02,Q:FarGolSip0001106,Q:FarGolPre01}.
We choose it here because a time-continuous description is more convenient when 
determining steady states. Our conclusions remain valid also for the usual iterative version.
We have verified this numerically for small qubit numbers $q<10$, where iterations with the unitary $G_w$ 
were alternated with finite time intervals of qubit decay. For larger qubit numbers
we expect the iterative algorithm to resemble a time-continuous process ever more closely.

\subsection{Description of the decay}

The oscillatory search behavior mentioned above is reminiscent of the Rabi oscillation in a
two-level system coupled to a resonant driving field. In the latter case
spontaneous decay damps the Rabi oscillation and
relaxes the driven two-level system to a steady state. Guided by this
analogy we now introduce spontaneous decay for the qubits, 
driving each qubit toward the state
$H|0\rangle=(1/\sqrt{2})(|0\rangle+|1\rangle)$ 
and the system as a whole toward the uniform superposition 
$|s\rangle$.

This choice of $|s\rangle$ as the ``ground state'' of the decay is essential.
In a loss-free search, starting with $|s\rangle$ as the initial state, the search Hamiltonian 
$\mathcal{H}_w$ drives a rotation in a two-dimensional subspace 
spanned by $|w\rangle$ and $|s\rangle$. The quantum state is then confined to this subspace. 
Dissipation or decoherence will in general lead to leakage of population out of 
this subspace. In our case, however, the decay prevents the quantum state from straying 
too far from $|s\rangle$ and thus from the subspace.

By our use of the word ``decay'' 
we mean to indicate that the process is unidirectional. We do not mean that $|s\rangle$,
the state the system decays into, is the lowest eigenstate of 
$\mathcal{H}_w$. In fact, all eigenvalues are zero, except the two highest ones, $1\pm 1/\sqrt{N}$,
corresponding to the eigenstates $|s\rangle\pm |w\rangle$ (not normalized). 
The choice to have the subspace $\{|s\rangle,|w\rangle\}$ at the highest energies 
rather than the lowest is arbitrary and unimportant for our results. 

It may seem unnatural to consider a process where the qubits decay to 
$H|0\rangle=(1/\sqrt{2})(|0\rangle+|1\rangle)$, the natural choice being perhaps
decay to $|0\rangle$ (or $|1\rangle$). 
However, our choice is equivalent to using
the ``natural'' decay to the ground state $|0\rangle$, 
if the entire quantum search is performed in a
basis transformed by $H^{\otimes q}$.
For convenience, in our simulation we choose to let the 
qubits decay to $H|0\rangle$ and to
perform the quantum search in the standard computational basis.

The decay is described by means of a master equation for the
density matrix $\rho$. We use a common model for spontaneous emission into a 
reservoir of modes at zero temperature. Thus using the Born-Markov approximation, 
and including also the coherent Hamiltonian evolution, 
it has the following form \cite{VisNie95},
\beq
\dot{\rho}=i\,[\rho,\mathcal{H}_w]+\frac{\Gamma}{2}\sum_{i=0}^{q-1}\left( 2\, c_i\,\rho\,
c_i^\dag-\, c_i^\dag c_i\,\rho-\,\rho\, c_i^\dag c_i\right)
\label{eq:master}
\eeq
where $\Gamma$ is a local (single-qubit) decay rate.
The $c_i$ are transition operators that describe
the coupling to a reservoir. They are chosen to be local 
lowering operators in the Hadamard-transformed basis,
\beq
c_i=I_2\otimes\cdots\otimes H d^- H \otimes\cdots\otimes I_2
\eeq
where the number of factors is $q$, and the single-bit lowering operator
$d^-= |0\rangle \langle 1|$ appears at position $i$. At all other positions we have the
$2\times 2$ identity matrix $I_2$. 

With the $c_i$ thus defined, each qubit decays to the state
$H |0\rangle=(1/\sqrt{2})(|0\rangle+|1\rangle)$. The collective ``ground
state'' of the decay process is therefore the uniform superposition
$|s\rangle$. This is a crucial difference with previous work by
Zhirov and Shepelyansky \cite{Q:ZhiShe06},
where the decay drove the system to the state $|0\rangle^{\otimes q}$.
In the latter case the buildup of population in $|w\rangle$ is not present in 
the steady state. In this paper we describe how this buildup does occur if 
the lowering operators $c_i$ are defined 
in a basis that is rotated with respect to the standard computational basis in which
the coherent search process is defined. As mentioned above, we could have chosen not to 
rotate the dissipation basis and to rotate the search basis instead.

\subsection{Integration of the master equation}

We can now simulate the quantum search in the presence of decay 
by integrating Eq.~(\ref{eq:master}).
An example is shown in Fig.~\ref{F:dampos}(a) for $q=6$ qubits. 
For comparison we also show the result without decay.
We notice that the oscillation of the population of the solution,
$\rho_{ww}$, is damped by the decay. The steady state value is clearly much 
higher than $2^{-q}\approx 0.016$, which would be the value if the population were
randomized entirely. A readout will thus reveal the
solution with increased probability. The readout will typically contain errors 
in some fraction of bits. The full solution can be retrieved after 
repeating this procedure a modest number 
of times, as we show in Sec.~\ref{sec:results}.

\section{Reduction of scaling by symmetry}
\label{sec:symmetry}

We now investigate how the steady state population $\rho_{ww}$
depends on the loss rate $\Gamma$ and how it scales with the number of
qubits $q$. The latter is particularly nontrivial, since simulating a
quantum computer on a classical one is usually inefficient. The number of entries in the
density matrix is $2^{2q}$ and quickly exhausts the resources of any
classical computer. Fortunately we can exploit a symmetry in the problem
to vastly reduce this number, and reduce the scaling to
polynomial in $q$.

This symmetry can be recognized in Fig.~\ref{F:dampos}(b), showing how 
the population $\rho_{xx}$ of any given basis state $|x\rangle$ 
depends only on the Hamming distance between $x$ and
$w$, defined as the number of bits in
which $x$ and $w$ differ. Inspection of the
full density matrix reveals a similar symmetry for the
off-diagonal elements, with $\rho_{xy}$ only a function
of the Hamming distances $d_{xw}$, $d_{yw}$, and $d_{xy}$.

To understand why this is the case we first perform a basis transformation
$S_w$ that for a given solution $w$ relabels the basis states as 
$S_w |x\rangle = |x'\rangle = | \textsc{XOR}(x,w)\rangle$, 
where the $\textsc{XOR}$ operation is to be taken bitwise.
Note that $S_w |w\rangle = |0\rangle$ so that the Hamming distances
become $d_{xw}\rightarrow d_{x0}$ and 
similarly $d_{yw}\rightarrow d_{y0}$. 
The Hamming distance $d_{xy}$ is invariant under $S_w$.

The ``shift'' operator $S_w$ transforms the search Hamiltonian into that for $w=0$:
\beq
S_w \mathcal{H}_w S_w^\dag  =  \mathcal{H}_0 = |0\rangle \langle 0| + |s\rangle \langle s|,
\eeq
while it leaves the decay terms in the master equation invariant. 
This can be seen by writing $S_w=I_2\otimes\cdots\otimes X\otimes\cdots\otimes I_2$,
where the number of factors is $q$ and the Pauli matrix 
$X=|0\rangle\langle 1|+|1\rangle\langle 0|$ appears in every position
$i$ where the corresponding bit value of $w$ is equal to 1. 
Defining $c=Hd^-H$ and $\tilde{c}=XcX$, one easily verifies that
$c^\dag c=\tilde{c}^\dag \tilde{c}$
and $c \rho_2 c^\dag=\tilde{c} \rho_2 \tilde{c}^\dag$,
for any arbitrary $2\times 2$ matrix $\rho_2$. From this it follows that the decay terms
in the master equation are unchanged by $S_w$.

From here on we consider the search problem with $w=0$ without loss of generality.
This problem is symmetric under bit swaps, i.e., under all possible
permutations of the qubits. This is immediately clear by inspection:
both $|0\rangle$ and $|s\rangle$ appearing in the
search Hamiltonian $\mathcal{H}_0$ are product states with all qubits in the same state.
The decay terms in the master equation consist of sums of identical terms for all qubits
and are therefore also invariant under qubit permutations.

When searching for the steady state
of the master equation, we can now restrict ourselves to density matrices that have 
bit swap symmetry. For a given matrix element $\rho_{xy}$, it is
convenient to introduce the bit pair counts $n_{00}$, $n_{01}$, $n_{10}$, and $n_{11}$,
such that $n_{00}$ counts the number of instances where corresponding
bits in $x$ and $y$ both have the value 0. The other bit-pair counts
$n_{01}$, $n_{10}$, and $n_{11}$ are defined similarly. Their relation
to $(d_{x0}, d_{y0}, d_{xy})$ is
\begin{align}
d_{x0} & = n_{10}+n_{11}\\
d_{y0} & = n_{01}+n_{11}\\
d_{xy} & = n_{01}+n_{10}\\
q & = n_{00}+n_{01}+n_{10}+n_{11}
\end{align}
If we perform a permutation of qubits, so that $x\rightarrow x'$ and $y\rightarrow y'$,
corresponding bits in $x$ and $y$ are permuted in the same way. Therefore any permutation
of qubits leaves the bit pair counts unchanged, so $\rho_{xy}=\rho_{x'y'}$.

\begin{figure}
\includegraphics[width=80mm]{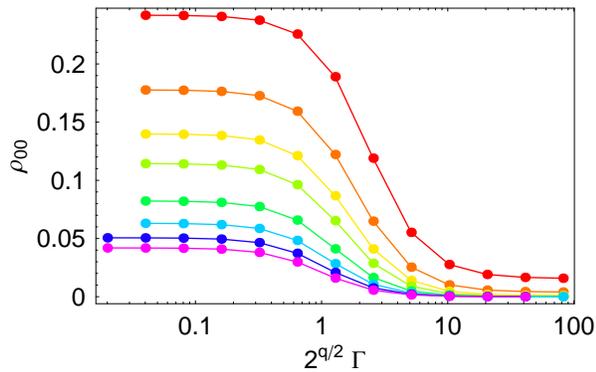}
\caption{(Color online) Steady state population in the solution, $\rho_{00}$, as a
function of the decay rate $\Gamma$.
The data sets are for different numbers of
qubits, from the top down: $q=6,8,10,12,16,20,24,28$.
The decay rate has been scaled by the frequency of the quantum
search oscillation, $\sim 2^{-q/2}$.}
\label{F:rhossvsgamma}
\end{figure}

Instead of labeling the density matrix elements
by index pairs, we now label them by $(n_{00}, n_{01}, n_{10}, n_{11})$:
\beq
    \rho_{xy}=\sigma_{n_{00}, n_{01}, n_{10}, n_{11}}.
\eeq
This is  a crucial result: whereas $x,y$ run from 0 to $2^q-1$, 
the bit pair counts (or Hamming distances) run only from 0 to $q$.
Since the bit pair counts sum to $q$, the number of different combinations of
bit pair counts is given by 
\beq
\left(
\begin{array}{c}
q+3 \\ 3
\end{array}
\right)
=\frac{1}{6}\,(q+1)(q+2)(q+3).
\eeq
This is a very modest number compared to the number of index pairs $(x,y)$, equal to
$2^{2q}$. The number of distinct entries in the density matrix is thus reduced
tremendously. More importantly, the exponential scaling with $q$ has
been replaced by a polynomial one. For example, for $q=36$, we have
$2^{72}\approx 4.7\times 10^{21}$ whereas  bit swap symmetry reduces this to 
only 9139 distinct matrix elements.

We now reexpress the master Eq.~(\ref{eq:master}) as the time evolution of the new
density matrix, $\dot{\sigma}$. The expression is somewhat lengthy but straightforward and is 
given in the appendix. 
Using this equation we can now either integrate it in time, or directly solve $\dot{\sigma}=0$
to obtain the steady state. It should be noted however that we have hereby given up the possibility 
to start the time integration with an arbitrary initial density matrix. The 
equation for $\dot{\sigma}$ implicitly assumes the symmetry described above. 
The usual initial condition, starting from the superposition $|s\rangle$ does have this 
symmetry and can therefore be simulated. 

If we solve directly for the steady state, we are not subject to this limitation. 
The decay process ensures that the density matrix acquires the appropriate symmetry over time. 
The usual initialization step is thus unnecessary. 
It is worth noting that the time needed to develop 
bit swap symmetry is the same as the time to reach the steady state. 
This is not surprising, considering that both are driven by the same decay process. 
Numerically, this was observed by choosing a randomly 
chosen, (pure,) initial state $\rho$, lacking bit swap symmetry. 
For obvious reasons this test is restricted to 
small numbers of qubits.

From here on we solve only the equation $\dot{\sigma}=0$, which is a system of linear 
equations with as many variables. More precisely, these linear equations are not independent. 
This is resolved by replacing one of the linear equations by the unit trace condition 
$\textrm{Tr}\, \rho =1$, which translates to 
\beq
  \textrm{Tr}\, \rho =\sum_{d_{x0}=0}^{q} \left(
  \begin{array}{c}
  q \\ d_{x0}
  \end{array} \right) \sigma_{q-d_{x0},0,0,d_{x0}} =1.
\eeq

\section{Numerical results for the steady state}
\label{sec:results}

In Fig.~\ref{F:rhossvsgamma} we show the resulting steady state
population in the solution, $\rho_{00}=\sigma_{q,0,0,0}$, for various numbers of qubits
$q$, and varying the loss rate $\Gamma$. The loss rate has been scaled
by the frequency of the loss-free quantum search oscillation
$~2^{-q/2}$. We clearly recognize two different regimes,
$2^{q/2}\,\Gamma\gg 1$ and $2^{q/2}\,\Gamma\ll 1$. 
We find that in both regimes the steady state
populations are independent of the decay rate $\Gamma$. There is a transition region 
around $2^{q/2}\,\Gamma\approx 1$ where the steady state population goes over 
from the low-$\Gamma$ value to the high-$\Gamma$ value. 
Not surprisingly, in the limit of strong
decay the steady state approaches $|s\rangle$, so that all
populations in the computational basis, $\rho_{xx}$, approach the value
$2^{-q}$.

The low-$\Gamma$ regime is thus reached when the {\em single-qubit} decay rate is 
well below the frequency of the loss-free search oscillation (the energy separation between 
the two highest eigenvalues of $\mathcal{H}_w$). This means that the probability that 
one particular qubit has decayed during one search period is small. However, 
the probability that at least one qubit decays during such a period can be high. 

On the basis of Fig.~\ref{F:rhossvsgamma} we can now estimate the total 
search time. We see from this figure that for 
$2^{q/2}\Gamma \lesssim 0.3$ we are essentially in the low $\Gamma$ limit. The time $T$
needed to reach the steady state is a few times $\Gamma^{-1}$. For the example of 
Fig.~\ref{F:dampos}(a) we have $2^{q/2}\Gamma = 0.24$ and $\Gamma T =3.6$ at $T=120$.
Thus the total search time is $T\approx 15\times 2^{q/2}$, roughly ten times longer than 
the value of $(\pi/2) 2^{q/2}$ for the loss-free case. 
The above estimate states essentially that ({\em i}) the search oscillation must be well 
underdamped, and ({\em ii}) we need to wait for a few damping times. Therefore 
we expect the mentioned factor of ten to be essentially constant, i.e. not to scale with $q$.

\begin{figure}
\includegraphics[width=80mm]{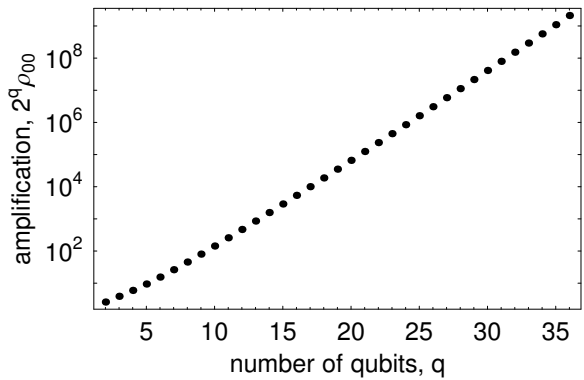}
\caption{Amplification factor 
$2^q \rho_{00}$, in the limit of weak decay,
versus the number of qubits $q$.}
\label{F:rhossplot}
\end{figure}

We now investigate the weak decay limit in some more detail. 
We choose, somewhat arbitrarily, $2^{q/2}\Gamma =0.005$ as representative of this low-$\Gamma$ limit. 
The steady state population of the solution, $\rho_{00}$, is then much larger
than the other populations. We also see that $\rho_{00}$ decreases with 
increasing $q$. A plot of $\rho_{00}$ vs. $1/q$ is remarkably well fitted by 
a straight line. However this relationship is purely 
heuristical and extrapolates to negative population for $q\gtrsim 120$.

We define $2^q \rho_{00}$ as an ``amplification
factor'', i.e. the factor by which $\rho_{00}$ is increased
compared to the fully random case. This is also the ratio of the 
$\rho_{00}$ values in the low-$\Gamma$ and the high-$\Gamma$ limits.
In Fig.~\ref{F:rhossplot} we show how the
amplification varies with $q$ in an approximately exponential fashion.

It is interesting to compare the weak decay case to a strongly
driven two-level system with spontaneous emission. The populations of the
ground and excited states then both approach the value 1/2. In the quantum
search the population $\rho_{00}$ is generally much smaller, except
for very small numbers of qubits. However, if we project the numerically obtained 
steady state on the subspace spanned by $\{|0\rangle,|s\rangle\}$, 
we do find equal populations in these two states.

\begin{figure}[tb]
\includegraphics[width=80mm]{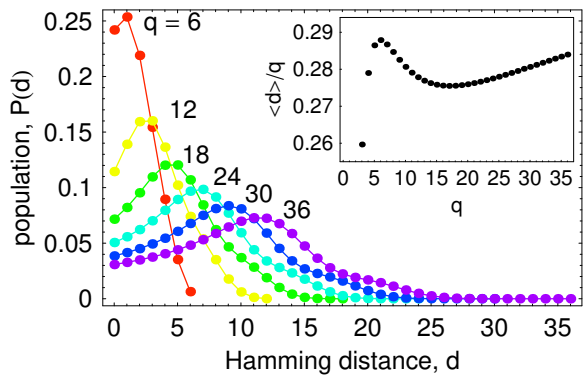}
\caption{(Color online) Distribution of steady state population over the Hamming
distance $d$ to the solution, in the weak decay limit.
The different data sets are for different numbers of
qubits, $q=6,12,18,24,30,36$. The insert shows the average Hamming distance
as a fraction of the number of qubits, versus $q$.}
\label{F:hddists}
\end{figure}

The observation that $\rho_{00}$ becomes ever smaller for
larger numbers of qubits may seem disappointing at first. However this is unjustly so.
We now investigate how the remaining population $1-\rho_{00}$ is distributed
and show that it is concentrated in states at small Hamming distance from the solution.

The diagonal elements of the density matrix are retrieved as
$\rho_{xx}=\sigma_{q-d_{x0},0,0,d_{x0}}$. Sorting these by
their Hamming distance $d_{x0}$, and
multiplying by their multiplicity (a binomial coefficient), we obtain the 
distribution of population over the Hamming distance $d$ to the solution. 
This is shown in Fig.~\ref{F:hddists}. For
small numbers of qubits the distribution is peaked at $d=0$. For larger
$q$ the peak shifts away from zero. This can be understood qualitatively 
because the multiplicity increases rapidly with $d$. 

For strong decay the mean of the distribution $\langle d\rangle =q/2$. 
A readout then yields a wrong bit value for
half the bits, on average.
For weak decay we find that the ``bit error rate'' $\langle d\rangle/q$ is again 
almost constant, with a value of 
$\xi\equiv\langle d\rangle/q\approx 0.28$, as can be seen in Fig.~\ref{F:hddists}.
The solution can now be
obtained by repeating the search a limited number of times using a majority vote rule to
decide about each individual bit value. We can estimate how many times $R$ we must
repeat the search and readout in order to increase the probability
to find an error-free solution to above a preset probability. 

The probability that the majority vote yields the wrong result 
for a particular bit is obtained by summing the binomial distribution, 
$\left(\begin{array}{c} R\\ n \end{array}\right)\xi^n (1-\xi)^{R-n}$,
for $n\ge (R+1)/2$ (for odd $R$). 
For large $R$ we can approximate the summation using standard techniques and 
obtain an upper bound for the single-bit error rate after majority voting,
\begin{equation}
\xi_R<\frac{1}{1-2\xi}\ \frac{(2\sqrt{\xi-\xi^2}\ )^{R+1}}{\sqrt{2 \pi R}}.
\label{eq:xiR}
\end{equation}
The probability of an error in at least one bit out of $q$ 
is then given by $1-(1-\xi_R)^q < q\,\xi_R$. 
For $q=29$ and $\xi=0.28$ we find that $q\,\xi_R<0.05\ (0.01)$ for $R\ge 41\ (53)$.

It is important to note the scaling behavior of $\xi_R$. From Eq.~(\ref{eq:xiR}) we see 
that $\xi_R$ decreases faster than exponential with $R$. Thus, if we demand that $q\,\xi_R$
drops below some preset error probability $\epsilon$, the number of 
repetitions will increase only as $R=O(\log (q/\epsilon))$.
The total search time then scales as $O(2^{q/2}\log (q/\epsilon))$, somewhat longer than for the 
commonly studied relaxation-free case ($2^{q/2}$) but still much shorter
than for a classical search ($2^q$).

\section{Conclusion}
\label{sec:conclusion}

In conclusion, we have shown that a modified, robust version of Grover's quantum search 
algorithm can be used if the qubits are subject to spontaneous decay. 
The search is conducted
in a Hadamard transformed basis and the single-qubit decay rate must be
smaller than the natural oscillation frequency of the search algorithm.

A symmetry allowed us to numerically analyze the problem for up to 
36 qubits on a standard desktop computer. This clearly opens up the possibility 
to investigate this problem for even larger number of 
qubits, at only polynomial costs. It remains an  open question
what will happen to the observed constant bit
error rate of 0.28 and the trend $\rho_{00}\sim 1/q$ in the limit $q\rightarrow\infty$. 
The latter must break down at around 100 qubits.
Possibly, an analytic approximation
in this limit could shed light on this issue.

An obvious extension of the present work is 
multiple-item searching for which we expect similarly robust behavior in the presence of relaxation. 
We can only speculate on the question whether similar modifications may be 
applied to other quantum algorithms.

\begin{acknowledgments}

We gratefully acknowledge M. Richter and H.B. van Linden van den Heuvell
for sharing computer facilities for the highest $q$ results, F. Terra and B. Rem 
for help in obtaining data.
This work is part of the research program
of the Stichting voor Fundamenteel Onderzoek van de Materie (Foundation
for the Fundamental Research on Matter) and was made possible by
financial support from the Nederlandse Organisatie voor Wetenschappelijk
Onderzoek (Netherlands Organization for the Advancement of Research).
\end{acknowledgments}

\appendix*
\section{Re-expression of the master equation}

Using the symmetry described above, we relabel the density matrix elements $\rho_{xy}$ using
the bit pair counts,  $\rho_{xy}=\sigma_{n_{00}, n_{01}, n_{10}, n_{11}}$.
The master equation can then be re-expressed as follows,
\begin{align}
& \dot{\sigma}_{n_{00}, n_{01}, n_{10}, n_{11}} = \nonumber \\
2^{-q}\,i\,& (\mathcal{R}-\mathcal{C}) + \nonumber \\
i \,& (\delta_{0,n_{01}+n_{11}} - \delta_{0,n_{10}+n_{11}})\,
    \sigma_{n_{00}, n_{01}, n_{10}, n_{11}} +   \nonumber \\
\frac{\Gamma }{4} \left[ \right.
    n_{00}\,& \left(
        \sigma_{ n_{00}-1, n_{01}, n_{10}, n_{11}+1} -
        \sigma_{n_{00}, n_{01}, n_{10}, n_{11}} \right)  +  \nonumber \\
    n_{11}\,& \left(
    \sigma_{n_{00}+1, n_{01}, n_{10}, n_{11}-1} -
    \sigma_{n_{00}, n_{01}, n_{10}, n_{11}} \right)  +  \nonumber \\
    n_{01}\,& \left(
    2\,\sigma_{n_{00}, n_{01}-1, n_{10}, n_{11}+1 } +
    2\,\sigma_{n_{00}+1, n_{01}-1, n_{10}, n_{11}} - \right. \nonumber \\
& \left. - \sigma_{n_{00}, n_{01}-1, n_{10}+1, n_{11}} -
        3\,\sigma_{n_{00}, n_{01}, n_{10}, n_{11}} \right)  + \nonumber \\
 n_{10} \,&\left(
    2\,\sigma_{n_{00}, n_{01}, n_{10}-1, n_{11} +1} +
    2\,\sigma_{n_{00}+1 , n_{01}, n_{10}-1, n_{11}} - \right. \nonumber \\
& \left. - \sigma_{n_{00}, n_{01}+1, n_{10}-1, n_{11}} -
    3\,\sigma_{n_{00}, n_{01}, n_{10}, n_{11}} \right) \left. \right]
\label{eq:sigmadot}
\end{align}
\newpage
where $\delta_{i,j}$ is the Kronecker delta and $\mathcal{R}$ and $\mathcal{C}$ are row and column sums, given by
\begin{multline}
\mathcal{R} =  \sum_{i_{00}=0}^{q-d_{x0}} \sum_{i_{11}=0}^{d_{x0}}
    \left( \begin{array}{c}
        d_{x0} \\ i_{11}
    \end{array} \right)
    \,\left( \begin{array}{c}
        q-d_{x0} \\ i_{00}
    \end{array} \right) \times \\
    \sigma_{i_{00}, q-d_{x0}-i_{00}, d_{x0}-i_{11}, i_{11}}
\end{multline}
\begin{multline}
\mathcal{C} =  \sum_{i_{00}=0}^{q-d_{y0}} \sum_{i_{11}=0}^{d_{y0}}
    \left( \begin{array}{c}
        d_{y0} \\ i_{11}
    \end{array} \right)
    \,\left( \begin{array}{c}
        q-d_{y0} \\ i_{00}
    \end{array} \right) \times \\
    \sigma_{i_{00}, d_{y0}-i_{11}, q-d_{y0}-i_{00}, i_{11}}
\end{multline}

\vspace*{80mm}


\end{document}